\documentclass[%
 reprint,
 aps,
 prx,
 amsmath,amssymb,
 superscriptaddress,
 floatfix,
]{revtex4-2}
\usepackage{hyperref}
\hypersetup{colorlinks=true,linkcolor=blue,citecolor=blue}
\usepackage[caption=false]{subfig}
\usepackage{graphicx}
\usepackage{dcolumn}
\usepackage{bm}
\usepackage[caption=false]{subfig}
\usepackage{xcolor}

\begin{document}

\preprint{APS/123-QED}

\title{Mitigation of Measurement-Induced State Transitions via a Fast-Load and Fast-Clear Readout}

\author{Wei-En Lin}
\thanks{These two authors contributed equally to this work.} 
\affiliation{Research Center for Critical Issues, Academia Sinica, Tainan, 711010, Taiwan}
\affiliation{Department of Physics, National Central University, Taoyuan, 320317, Taiwan}

\author{Li-Chieh Hsiao}
\thanks{These two authors contributed equally to this work.}
\affiliation{Research Center for Critical Issues, Academia Sinica, Tainan, 711010, Taiwan}
\affiliation{Department of Physics, National Cheng Kung University, Tainan, 701401, Taiwan}

\author{Chen-Hsun Ma}
\affiliation{Research Center for Critical Issues, Academia Sinica, Tainan, 711010, Taiwan}
\affiliation{Department of Physics, National Taiwan University, Taipei, 106319, Taiwan}

\author{Erh-Hsiang Yeh}
\affiliation{Department of Physics, National Central University, Taoyuan, 320317, Taiwan}

\author{Wei-Lun Peng}
\affiliation{Department of Physics, National Changhua University of Education, Changhua, 500207, Taiwan}

\author{Hsi-Sheng Goan}
\affiliation{Department of Physics, National Taiwan University, Taipei, 106319, Taiwan}
\affiliation{Center for Quantum Science and Engineering, National Taiwan University, Taipei, 106319, Taiwan}
\affiliation{Physics Division, National Center for Theoretical Sciences, Taipei, 106319, Taiwan}

\author{Cen-Shawn Wu}
\affiliation{Research Center for Critical Issues, Academia Sinica, Tainan, 711010, Taiwan}
\affiliation{Department of Physics, National Changhua University of Education, Changhua, 500207, Taiwan}

\author{Yueh-Nan Chen}
\affiliation{Department of Physics, National Cheng Kung University, Tainan, 701401, Taiwan}
\affiliation{Physics Division, National Center for Theoretical Sciences, Taipei, 106319, Taiwan}
\affiliation{Center for Quantum Frontiers of Research and Technology, National Cheng Kung University, Tainan, 701401, Taiwan}

\author{Yung-Fu Chen}\email{yfuchen@cc.ncu.edu.tw}
\affiliation{Department of Physics, National Central University, Taoyuan, 320317, Taiwan}
\affiliation{Center of High Energy and High Field Physics, National Central University, Taoyuan, 320317, Taiwan}
\affiliation{Quantum Technology Center, National Central University, Taoyuan, 320317, Taiwan}
\affiliation{Taiwan Semiconductor Research Institute, National Institutes of Applied Research, Hsinchu, 300091, Taiwan}
\affiliation{National Center for Excellence in Quantum Information Science and Engineering, National Tsing Hua University, Hsinchu, 300044, Taiwan }

\author{Chung-Ting Ke}\email{ctke@gate.sinica.edu.tw}
\affiliation{Research Center for Critical Issues, Academia Sinica, Tainan, 711010, Taiwan}
\affiliation{Institute of Physics, Academia Sinica, Taipei, 115201, Taiwan}

\author{Chii-Dong Chen}
\affiliation{Research Center for Critical Issues, Academia Sinica, Tainan, 711010, Taiwan}
\affiliation{Institute of Physics, Academia Sinica, Taipei, 115201, Taiwan}

\begin{abstract}
High-fidelity and rapid qubit readout is essential for superconducting quantum processors, typically realized through the quantum non-demolition (QND) dispersive interaction within a qubit-resonator architecture. However, the achievable readout speed and fidelity are fundamentally limited by measurement-induced state transitions (MIST). For a transmon qubit, MIST is highly sensitive to the offset charge $n_g$ due to the charge dispersion of its higher-lying energy levels. In this work, we systematically investigate $n_g$-dependent MIST dynamics governed by the diabaticity and symmetry of pulse shaping within a charge-sensitive
transmon architecture. We engineer fast-load and fast-clear pulses that effectively suppress resonator photon
overshoots, thereby demonstrating a highly practical strategy to mitigate MIST without requiring complex waveforms or real-time feedback. Utilizing active gate-voltage control and rapid feedback, the measurement-induced transition probability is precisely mapped against $n_g$ and the steady-state resonator photon number, exhibiting strong agreement with numerical Floquet branch analysis. Ultimately, we evaluate the $n_g$-averaged total error probabilities for both readout and post-readout stages, verifying that a straightforward three-step pulse scheme consistently minimizes overall readout errors. Within the framework of large-scale superconducting quantum processors, this practical, hardware-free approach inherently offers a better trade-off between the readout signal-to-noise ratio and QND preservation.

\end{abstract}

\maketitle


\section{Introduction}

Superconducting circuits incorporating transmon qubits~\cite{Blais:2004pra, Wallraff:2004nature, Koch:2007pra} have emerged as a preeminent platform for quantum information processing~\cite{Blais:2021rmp, Arute:2019nature, Krinner:2022nature, Google:2023nature}. In particular, fast and high-fidelity quantum state measurement is a prerequisite for implementing quantum error correction and feedback control toward fault-tolerant quantum computing~\cite{Fowler:2012pra, Google:2023nature, Krinner:2022nature}. In scalable architectures such as the surface code, syndrome extraction inherently demands millions of such rapid measurements~\cite{Fowler:2012pra}. Typically, a quantum non-demolition (QND) readout is achieved through the dispersive interaction between a qubit and a readout resonator~\cite{Blais:2021rmp}. To mitigate qubit relaxation errors during the readout, increasing the measurement photon number offers the most direct approach to enhance the signal-to-noise ratio (SNR), thereby enabling a shorter measurement time~\cite{Gambetta:2008pra, Jeffrey:2014prl, Walter:2017prapplied}. However, this strategy is fundamentally limited by the breakdown of the QND nature due to measurement-induced state transitions (MIST), indicating that the qubit is sometimes even excited out of the computational subspace \cite{Sank:2016prl}. Therefore, MIST significantly degrades readout fidelity and strictly constrains the measurement speed. Crucially, this vulnerability to measurement-induced leakage represents a universal challenge not only for transmons but for any driven open quantum system with higher-lying energy levels, including increasingly prominent fluxonium architectures~\cite{wjdb_4814,zwanenburg2026,chapple2026}. The underlying mechanism of MIST arises from the complex energy ladder structure of the coupled transmon-resonator system. Specifically, previous investigations using the Jaynes-Cummings model beyond the rotating wave approximation have successfully predicted the high-energy multiphoton resonances that occur at specific photon numbers well below both the ionization threshold and the conventional critical photon number~\cite{Sank:2016prl, Khezri:2023prapplied}. While these approaches successfully elucidate how such spurious resonances compromise the QND nature even under weak readout drives, determining the precise onset of ionization requires a more rigorous treatment. Recent work has established a comprehensive framework for transmon ionization~\cite{lescanne2019escape,Verney2019structural,Dumas:2024prx,shillito2022dynamics}, demonstrating that the existence of an ionization threshold signifies the failure of the Kerr nonlinearity approximation. Consequently, a non-perturbative model incorporating the exact full cosine potential with full matrix elements is required to accurately predict the onset of ionization in strongly driven systems.

In principle, even in the transmon regime defined by $E_J/E_C \gtrsim 50$, where $E_J$ and $E_C$ denote the Josephson energy and charging energy, respectively, MIST mechanisms are inherently sensitive to the offset charge $n_g$ due to the significant charge dispersion associated with the higher energy levels of the transmon \cite{Khezri:2023prapplied,Kurilovich:2025arxiv,Cohen2023Reminiscence,fechant2025offset}. 
While most studies on MIST assume an average over $n_g$ \cite{Sank:2016prl, Khezri:2023prapplied,Dumas:2024prx}, the explicit $n_g$-dependent dynamics of MIST have recently been revealed in Ref.~\cite{fechant2025offset}. The $n_g$-dependence of ionization dynamics via Landau-Zener experiments has been investigated in a typical transmon \cite{Wang2025probing}; building upon these insights into strong-drive dynamics, recent strategies have focused on spectral engineering to mitigate such transitions. Notably, implementing a highly detuned readout resonator relative to the qubit frequency effectively avoids spurious multi-excitation resonances, as demonstrated experimentally and corroborated by Floquet-mode simulations~\cite{Kurilovich:2025arxiv}; the residual transition mechanisms remaining in this regime have been characterized separately~\cite{Connolly:2025arxiv}. The time drift of $n_g$ resulting in unstable MIST is observed \cite{Hirasaki2024dynamics,Wang2025probing} and alleviated by adding an inductive shunt to the transmon \cite{IST2026measurement}.

The transient response of the readout signals is fundamentally governed by the finite bandwidth of the readout resonator. A slow photon ring-up inherently degrades the achievable SNR within a given readout window, while an excessive ring-down time limits the speed of active feedback control. To overcome these device constraints and maximize readout speed, implementing fast-load and fast-clear pulses has become a widely adopted and essential strategy \cite{Walter:2017prapplied,McClure:2016prapplied}. In this work, we demonstrate the effectiveness of the fast-load and fast-clear scheme in mitigating MIST, while systematically investigating $n_g$-dependent MIST dynamics governed by the diabaticity and symmetry of pulse shaping. Utilizing a branch analysis under an established semi-classical model, we numerically characterize the time-dependent behavior of dressed-state hybridization for typical pulse shapes. To this end, we synthesize fast-load and fast-clear waveforms that suppress transient photon overshoots, thereby offering a highly practical strategy to mitigate MIST without requiring complex waveform engineering or real-time feedback.

To experimentally validate these underlying mechanisms, we employ fixed-frequency floating transmon devices operating in the regime of $E_J/E_C \approx 34$--$40$, which are inherently susceptible to charge dispersion. Utilizing active gate-voltage control and rapid feedback acquisition to stabilize $n_g$ against temporal drift, we precisely map the measurement-induced transition probability $P_m$ as a function of both $n_g$ and the resonator photon number for the different pulse profiles.  Through the use of consecutive readout pulses, our measurement protocol successfully resolves the MIST dynamics arising from both the photon ring-up and the combined ring-up and ring-down backaction, yielding good agreement with numerical simulations. Ultimately, we evaluate the $n_g$-averaged total error probabilities for both readout and post-readout stages, which combine $P_m$ and overlap error to characterize the overall readout error as a function of the resonator photon number. Notably, we verify that the straightforward three-step pulse scheme consistently minimizes these errors among all investigated pulse profiles, providing a robust, hardware-efficient solution to effectively suppress overall readout backaction across the entire $n_g$ range without compromising the readout speed.

\section{Physical Model and Diabatic Mechanism}

Increasing the readout strength exacerbates MIST in a transmon when the driven dressed eigenstates undergo severe hybridization at avoided crossings. This process is governed by Landau-Zener dynamics at avoided crossings \cite{Ivakhnenko2023,Grifoni1998} and induces non-QNDness as well as significant readout infidelity \cite{shillito2022dynamics,Dumas:2024prx,Wang2025probing}. The situation becomes considerably more severe when considering the time drift of $n_g$ as it leads to substantial uncertainty in the calibrated readout fidelity. However, $P_{m}$ can be effectively mitigated by optimizing the diabaticity of the trajectory of resonator photon number $n(t)$ during the readout stage.

In this section, we employ input-output theory to calculate the time evolutions of the resonator photon number for different pulse shapes and determine the instantaneous slope near the cross resonant points. Using the \texttt{QuantumToolbox.jl} package \cite{QuantumToolbox.jl2025} for numerical implementation, we then perform Floquet analysis to evaluate the resulting MIST for each trajectory.  This methodology allows us to clarify the relationship between the instantaneous photon evolution rate and the transition probability, which highlights the importance of diabaticity in suppressing MIST.

\subsection{Semiclassical driven transmon model}

We consider a transmon-resonator system with coupling strength $g$, where the driving field applied to the resonator is treated as a classical coherent drive. Under this semiclassical approximation, and neglecting quantum fluctuations, the driven transmon Hamiltonian is expressed as \cite{Dumas:2024prx}
\begin{equation}
\hat{H}(t) = \hat{H}_t + \varepsilon_t(t)\cos(\omega_dt)(\hat{n}_t - n_g),
\label{eq_full_H}
\end{equation}
where $\hat{H}_t$ represents the static transmon Hamiltonian characterized by a full cosine potential
\begin{equation}
\hat{H}_t = 4E_C(\hat{n}_t - n_g)^2 - E_J\cos(\hat{\varphi}_t).
\label{eq_H_t}
\end{equation}
In these expressions, $\hat{n}_t$ and $\hat{\varphi}_t$ denote the canonically conjugate charge and phase operators of the transmon, respectively. The drive frequency $\omega_d$ is set close to the bare resonator frequency $\omega_r$. The effective driving amplitude, which originates from the capacitive coupling between the transmon and the resonator field, is defined as $\varepsilon_t(t) = 2g\sqrt{n(t)}$, where $n(t)$ represents the instantaneous resonator photon number derived from the resonator coherent state amplitude. 

\subsection{Floquet branch analysis}
To understand the origin of MIST, we analyze the semiclassical driven transmon model in the Floquet picture. We treat the system Hamiltonian as locally periodic since $\kappa\ll\omega_d$, where $\kappa$ indicates the total photon decay rate of the resonator \cite{Dumas:2024prx}. This approximation allows us to calculate an instantaneous Floquet spectrum by assuming a constant amplitude $\varepsilon_t$ \cite{breuer1989quantum}. For a specific $\varepsilon_t$, we solve the general eigenvalue equation $\hat{U}(t+T,t)|\phi(t)\rangle=e^{-i\epsilon T}|\phi(t)\rangle$ by diagonalizing the time evolution propagator $\hat{U}(t+T,t)$ defined over one drive period $T=2\pi/\omega_d$. Through the Floquet branch analysis, we extract the specific Floquet modes $|\phi_i[\varepsilon_t](t)\rangle$ (denoted simply as $|\phi_i[\varepsilon_t]\rangle$) and their corresponding quasienergies $\epsilon_i[\varepsilon_t]$. The branch index $i$ is rigorously established in the limit of zero drive amplitude, where the Floquet modes reduce to the bare transmon eigenstates $|\phi_i[0]\rangle=|i\rangle$. Under this scheme, multi-photon resonances at specific photon numbers manifest as avoided crossings in the quasienergy spectrum, indicating significant hybridization between distinct Floquet modes. For each branch $B_i$, the lowest photon number at which these resonant structures occur is defined as the critical threshold, $n_{i,\text{cri}}$. Pinpointing these critical values offers a rigorous diagnostic to determine where MIST is initiated, eventually leading to transmon ionization.
Crucially, rather than employing the adiabatically recursive tracking procedure used in Ref.~\cite{Dumas:2024prx,fechant2025offset}, we label each Floquet mode with branch $B_i$ by maximizing the overlap $\vert\langle\phi[\varepsilon_t]\vert\phi_i[0]\rangle\vert$. Because the driving pulse manipulation in this work operates in a relatively diabatic regime, our simulation within this framework yields a much better match with the measured data.  

When the system traverses these avoided crossings of dressed states, the resulting diabatic behavior can be modeled as a Landau-Zener transition \cite{Drese1999,Ikeda2022,Dumas:2024prx,Wang2025probing}. According to the principles of Landau-Zener dynamics, the instantaneous evolution rate of the resonator photon number $\Gamma_{i,r} = dn(t)/dt|_{n_{i,\text{cri}}}$ is proportional to the Landau-Zener speed
\begin{equation}
v = \Gamma_{i,r}\sqrt{2\Delta_{\text{ac}}\frac{d^2\epsilon_i(n)}{dn^2}\Big|_{n_{i,\text{cri}}}},
\label{eq_PLZ_speed}
\end{equation}
where $\epsilon_i(n)$ represents the Floquet quasienergies as a function of $n$, and $\Delta_{\text{ac}}$ represents the quasienergy gap of the Floquet modes. Thus, the diabatic transition probability at the avoided crossing point is well characterized by the Landau-Zener formula
\begin{equation}
P_{\text{LZ}} = \exp \left( \frac{-\pi \Delta^2_{\text{ac}}}{2v} \right).
\label{eq_PLZ}
\end{equation}
Here, a higher speed $v$ drives $P_{\text{LZ}}$ toward unity. Due to the strong hybridization at the avoided crossing, the physical character of the adiabatic branch drastically changes, evolving from the initial bare state into a highly excited state and thereby causing MIST. Therefore, a large $P_{\text{LZ}}$ implies a highly diabatic passage, meaning the system tunnels across the avoided crossing gap with high probability to the other adiabatic branch, effectively following the diabatic trajectory instead of the adiabatic process, which physically preserves its original bare state. While Eq.~(\ref{eq_PLZ}) provides an analytical expression conditioned strictly on a constant $\Gamma_{i,r}$ near the avoided crossing, the explicit time evolution of branches is still required to account for the dynamics of complex photon trajectories with a time-varying $\Gamma_{i,r}$ and consecutive crossings.

\begin{figure*}[t]  
\centering
\includegraphics[width=\textwidth]{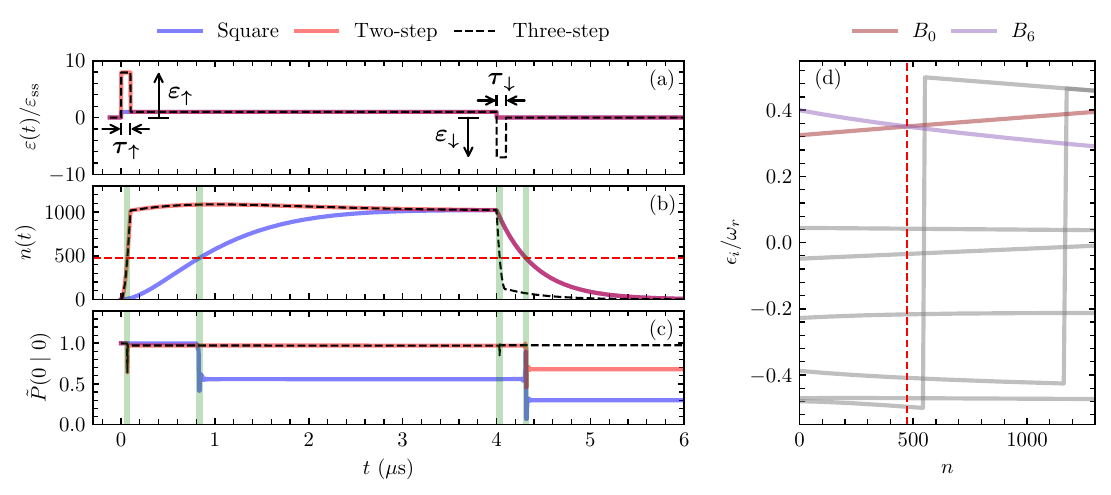} 
\caption{Floquet analysis of various pulse shapes. (a) Pulse amplitudes $\varepsilon(t)$ normalized by the steady-state amplitude $\varepsilon_{\text{ss}}$, (b) the corresponding photon populations $n(t)$, and (c) time-dependent branch probabilities $\tilde P(0\mid 0)$ evaluated via Eq.~(\ref{branch_P}), for square (blue), two-step (red), and three-step (black dashed) pulses, all initialized at $t=0$. ($\tau_{\text{ss}}+\tau_\uparrow=4\,\mu\text{s}$ and $\tau_\uparrow=\tau_\downarrow=100\,\text{ns}$). (d) Normalized quasienergies $\epsilon_i / \omega_r$ as a function of the instantaneous resonator photon number  $n$, with brown and purple curves denoting branches $B_0$ and $B_6$, respectively. The red dashed lines in (b) and (d) mark the critical crossing point, located at $n = n_{0,\text{cri}} = 474$, while the green-shaded regions in (b) and (c) highlight the instants when the photon trajectories traverse this critical threshold. }
\label{fig1}
\end{figure*}

To describe the full time-dependent dynamics starting from the initial state $\vert\Psi_i(0)\rangle=\vert i\rangle$, we solve the time-dependent Schrödinger equation governed by the Hamiltonian in Eq.~(\ref{eq_full_H}) to obtain the time-evolved state $\vert\Psi_i(t)\rangle$. Consequently, the population redistribution of the time-evolved state can be expressed as 
\begin{equation}
\tilde P(j\mid i)(t)=\vert\langle\phi_j[\varepsilon_t(t)]\vert\Psi_i(t)\rangle\vert^2, 
\label{branch_P}
\end{equation} 
which denotes the instantaneous transition probability from the prepared state $\vert i\rangle$ to branch $B_j$. Accordingly, the measurement-induced transition probability is time-averaged over the readout window as
\begin{equation}
P^i_m(t_1,t_2)=\frac{1}{t_2-t_1}\sum_{j\neq i}\int_{t_1}^{t_2}\tilde P(j\mid i)(t)dt,
\label{outlier_cal}
\end{equation}
where $t_1$ and $t_2$ denote the start and end times of the readout process, respectively. The probability $P_{m}^{i}$ characterizes the observed MIST during the readout, which is significantly affected by both the photon ring-up and ring-down processes. Specifically, when $n(t)$ approaches $n_{i,\text{cri}}$ during the ring-up process, the population of the associated branches undergoes non-adiabatic splitting governed by the Landau-Zener transition. Subsequently, Landau-Zener-Stückelberg (LZS) interference naturally occurs as the system traverses the resonance again during the ring-down process \cite{Shevchenko2010}. Because MIST is ultimately determined by the population transfers at these crossing points, this perspective implies that the suppression of $P^i_m$ can be achieved by maximizing diabaticity. We implement this strategy by engineering the passage speeds across the resonance during the respective transient stages via fast-load and fast-clear pulse shaping, rather than through active phase synchronization of the interference.

\subsection{Fast-load pulse and diabatic passage}\label{section_fast_load}

According to the prediction of Eq.~(\ref{eq_PLZ}), a rapid photon ring-up suppresses MIST, while the shortened transient response of the readout signal simultaneously enhances the readout SNR. To investigate the diabaticity associated with a large $\Gamma_{i,r}$ during the photon ring-up, we employ a fast-load scheme via a piecewise constant pulse with parameters $\{\varepsilon_\uparrow, \tau_\uparrow\}$ for the initial ring-up and $\{\varepsilon_{\text{ss}}, \tau_{\text{ss}}\}$ for the subsequent steady-state (ss) stage. Here, the symbols $\varepsilon$ and $\tau$ represent the drive amplitude and duration of each segment, respectively. Within the framework of a notch-type readout resonator coupled to a transmon qubit and a transmission line, the time evolution of the resonator field is determined segment by segment using input-output theory.  Under a constant drive of amplitude $\varepsilon=\sqrt{\frac{P_\text{in}}{\hbar\omega_d}}$ 
with an input power $P_{\text{in}}$, the general solutions for the resonator fields corresponding to the qubit states $|0\rangle$ and $|1\rangle$ are expressed as
\begin{equation}
\alpha_{\pm}(t)=\left(\alpha_{\pm}(0)-\mathcal{A}_{\pm}\right)e^{-[\kappa/2+i(\Delta_{r}\pm\chi)]t}+\mathcal{A}_{\pm},
\label{alpha_ge}
\end{equation}
where the minus and plus signs $\pm$ correspond to $|0\rangle$ and $|1\rangle$, respectively, and the steady-state amplitude as a function of $\varepsilon$ is given by
\begin{equation}
\mathcal{A}_{\pm}(\varepsilon) = \frac{-\varepsilon \sqrt{\kappa_c/2}}{\kappa/2+i(\Delta_r \pm \chi) }.
\end{equation}
In these expressions, $\kappa = \kappa_i + \kappa_c$ represents the total photon decay rate, comprising the intrinsic loss $\kappa_i$ and the external coupling $\kappa_c$. The term $\chi$ represents the dispersive shift, and the resonator detuning is defined by $\Delta_r = \widetilde{\omega}_r - \omega_d$, where $\widetilde{\omega}_r$ represents the average of the two dispersively shifted resonator frequencies corresponding to the qubit states. A detailed derivation of Eq.~(\ref{alpha_ge}) is provided in Appendix~\ref{appendix_io_theory}. Crucially, the final value of $\alpha_{\pm}$ from the first segment determines the initial condition $\alpha_{\pm}(0)$ for the subsequent evolution.

The resonator transient response in the readout integration window degrades state distinguishability during the ring-up process, consequently exacerbating readout overlap errors \cite{Wei2026}. To mitigate this, the fast-load scheme (such as the two-step pulse) enables the readout signal to reach its steady state much faster than under a standard square pulse drive at the same steady-state photon number, $n_{\text{ss}} \equiv \max_{i \in \{+,-\}} n_i(t\rightarrow\infty)$, where $n_{i}(t) = \left\vert \alpha_{i}(t) \right\vert^{2}$ and $i$ denotes the label of the qubit states.

Despite the potential issue of transient photon overshoot, even an uncalibrated two-step readout accelerates the resonator ring-up, thereby reducing readout overlap errors. To suppress the transient photon overshoot associated with the rapid ring-up, we impose the condition $\alpha_{\pm}(\tau_\uparrow) = \mathcal{A}_{\pm}(\varepsilon_{\text{ss}})$. Specifically, the drive amplitude $\varepsilon_{\uparrow}$ is chosen such that the resonator field at the end of the first segment ($t = \tau_\uparrow$) precisely matches the target steady-state resonator field of the second segment. Because this matched state precisely coincides with the steady-state fixed point under the readout drive $\varepsilon_{\text{ss}}$, the system immediately settles into its stationary state upon switching, thereby eliminating any subsequent transient photon overshoot. From the analytical expressions in Eq.~(\ref{alpha_ge}), the required drive amplitudes conditioned on the qubit state are directly determined as
\begin{equation}
\varepsilon_{\uparrow,\pm} = \frac{\varepsilon_{\text{ss}}}{1 - e^{-[\kappa/2 + i(\Delta_r \pm \chi)] \tau_\uparrow}} .
\end{equation}
While this conditional solution provides the optimal ring-up for each state, a more practical implementation that avoids the complexities of real-time state-dependent feedback is to intentionally bypass the complex phase correction by setting $\Delta_r = 0$ and $\left|\alpha_{\pm}(\tau_\uparrow)\right| = \left|\mathcal{A}_{\pm}(\varepsilon_{\text{ss}})\right|$. The required drive amplitudes for both states are equalized, yielding the state-independent amplitude
\begin{equation}
\varepsilon_{\uparrow} = \left| \frac{\varepsilon_{\text{ss}}}{1 - e^{-(\kappa/2 + i\chi) \tau_\uparrow}} \right|.
\label{cal_two_step}
\end{equation}
This formulation enables the adjustment of the ring-up speed while maintaining a constant $n_{\text{ss}}$. Despite lacking state-dependent phase compensation, it still effectively suppresses the large transient photon overshoots characteristic of uncalibrated two-step pulses. By establishing this precise control scheme, the strategic adjustment of $\varepsilon_{\uparrow}$ and $\tau_\uparrow$ allows for the deliberate shaping of the resonator photon trajectory $n(t)$ across critical avoided crossings. Recalling the Landau-Zener dynamics in Eqs.~(\ref{eq_PLZ_speed}) and (\ref{eq_PLZ}), a sufficiently high $\Gamma_{i,r}$ minimizes $P_{m}^{i}$.  Consequently, the synthesized pulse profile governs the competition between adiabatic hybridization and diabatic passage at these crossing points. Crucially, the qubit remains susceptible to MIST during ring-down, as the second traversal of $n_{i,\text{cri}}$ further contributes to the accumulation of the overall $P_m$, potentially exacerbated by the slower and more adiabatic nature of the ring-down passage \cite{fechant2025offset,Wang2025probing}. To alleviate this effect, a three-step pulse is employed to enforce a near-symmetric, rapid ring-up and ring-down of $n(t)$, effectively accelerating the ring-down through a similar treatment as the two-step pulse. The active ring-down segment $\{\varepsilon_{\downarrow}, \tau_{\downarrow}\}$ is applied after a duration of $\tau_{\uparrow} + \tau_{\text{ss}}$. A detailed derivation of this scheme is provided in Appendix~\ref{appendix_three_step}. 

To demonstrate this mechanism, we investigate the MIST behavior across three distinct pulse profiles using a qubit-resonator system with parameters of Q1 provided in Table~\ref{Table1}. To ensure the system is driven into the active MIST regime, we apply the readout pulses at $\Delta_r=0$ with a fixed steady-state power of $P_{\text{in}}=-113.3\,\text{dBm}$, which intentionally yields $n_{\text{ss}}>n_{0,\text{cri}}$. The pulse waveforms are illustrated in Fig.~\ref{fig1}(a), characterized by $\tau_{\uparrow}=100$\,ns for the two-step pulse and $\tau_{\uparrow}=\tau_{\downarrow}=100$\,ns for the three-step pulse.  As depicted in Fig.~\ref{fig1}(b), $n_{+}(t)=n_{-}(t)\equiv n(t)$ due to the condition $\Delta_r=0$. As expected, the state-independent two-step and three-step pulses reach the target $n_\text{ss}$ without the detrimental photon overshoot. To ensure a fair comparison of hybridization effects relative to the diabaticity of the passage, we select the parameter sets $\{\varepsilon_\uparrow, \tau_\uparrow\}$ that maintain a constant $n_{\text{ss}}$.

\begin{table}[h!]
\caption{Device parameters.}
\label{tab_params}
\begin{ruledtabular}
\begin{tabular}{lccc}
Parameter & Q1 & Q2 \\
\hline
Bare resonator frequency $\omega_{r}/2\pi$ (GHz) & 4.912 & 5.113 \\
 Qubit transition frequency  $\omega_{01}/2\pi$ (GHz) & 3.078 & 3.356 \\
 Qubit-resonator coupling strength  $g/2\pi$ (MHz) & 26.3 & 27.9 \\
 Dispersive shift    $\chi/2\pi$ (MHz) & -0.074 & -0.083 \\
 Resonator linewidth $\kappa/2\pi$ (MHz) & 0.392 & 0.352 \\
 Anharmonicity  $\alpha/2\pi$ (GHz) & -0.240 & -0.236 \\
 Josephson energy $E_J/2\pi$ (GHz) & 6.762 & 7.944 \\
 Charging energy $E_C/2\pi$ (GHz) & 0.201 & 0.201 \\
   $E_J/E_C$ & 33.64 & 39.5 \\
 Critical photon number $n_{\rm{cri}}=\frac{(\omega_{01}-\omega_{r})^2}{4g^2}$ & 1216 & 991 \\
\end{tabular}
\end{ruledtabular}
\label{Table1}
\end{table}

In Fig.~\ref{fig1}(d), we label each branch index $i$ with its corresponding quasienergy $\epsilon_i$ obtained via numerical Floquet branch analysis at $n_g=0.4$. A representative branch swapping between $B_0$ and $B_6$ is illustrated. Since the avoided crossing occurs at $n = n_{0,\text{cri}}=474$, we evaluate the time-dependent branch probability $\tilde P(0\mid 0)$ via Eq.~(\ref{branch_P}) initialized in the $\vert0\rangle$ state. As the photon trajectories corresponding to the three pulse profiles traverse the critical point $n_{0,\text{cri}}$ during both the ring-up and ring-down processes, the population of $B_0$ depletes due to state hybridization. As illustrated in Figs.~\ref{fig1}(b) and (c), a lower $\Gamma_{i,r}$ clearly exacerbates the severity of MIST.  Both the two-step and three-step pulses simultaneously minimize the resonator transient response while suppressing MIST. Furthermore, the three-step configuration offers enhanced mitigation specifically for the ring-down process. These synergistic improvements demonstrate that maximizing readout diabaticity is achievable via straightforward pulse shaping.  This simple pulse design minimizes ring-up and ring-down durations without compromising the readout speed. While the preceding example considers a single offset charge, the critical photon threshold inherently becomes $n_g$-dependent due to the charge-dependence of higher-lying transmon levels, and is thus expressed as $n_{i,\text{cri}}(n_g)$. By defining the $n_g$-dependent $\Gamma_{i,r}$ for each transmon state $\vert i \rangle$ as $\Gamma_{i,r}(n_g) = \frac{dn}{dt} \Big|_{n = n_{i,\text{cri}}(n_g)}$, it becomes intuitive that maintaining high diabaticity ensures robust overall suppression of MIST over the entire $n_g$ range. This motivates us to investigate MIST structures as a function of the readout photon number and offset charge, focusing on commonly used pulse shapes optimized for fast readout and straightforward implementation.

\section{Experimental Implementation}

\subsection{Circuit architecture and sample parameters}
As illustrated in Fig.~\ref{fig:FIG2}(a), the chip consists of two floating, Al-based, single-junction transmon qubits, labeled Q1 and Q2. Each qubit is capacitively coupled to a dedicated $\lambda/4$ readout resonator. The resonators are inductively coupled to a common transmission line through which qubit control pulses, readout pulses, and voltage biases are all delivered via a bias tee. Q1 and Q2 are charge-sensitive transmons that are dispersively coupled to their respective readout resonators, as detailed in Table~\ref{Table1}. The relaxation times for Q1 and Q2 are measured as $T_1 = 293\,\mu\text{s}$ and $135\,\mu\text{s}$, respectively. At the charge sweet spot $n_g = 0$, the decoherence times are found to be $T_2 = 84\,\mu\text{s}$ and $25\,\mu\text{s}$.
The device exhibits low effective qubit temperatures of $T_q=18.46\,$mK and $16.86\,$mK, both close to the bath temperature of $\sim17\,$mK, indicating a well-isolated experimental environment. Details of the wiring configuration and the raw data used for the $T_q$ characterization are provided in Appendix~\ref{Experiment Setup}.

\subsection{Experimental method}\label{sec:Exp_meth}

\begin{figure}
\centering
\includegraphics[width=\columnwidth]{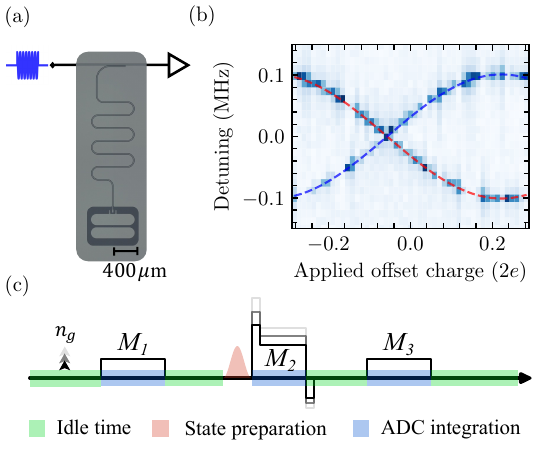}
\caption{Device architecture and MIST characterization framework. (a) Optical micrograph of the device comprising a floating transmon coupled to its dedicated readout resonator, which is inductively coupled to a common transmission line. (b) Detuning frequencies extracted from the Fourier transform of Ramsey measurements at 3.077\,GHz, showing the even- and odd-parity qubit frequencies as a function of applied offset charge for Q1. (c) Schematic of the MIST characterization protocol, incorporating three distinct analog-to-digital converter (ADC) signal integration windows.  After the Ramsey-based calibration of $n_g$ and stabilization at the target bias point, the $M_1$ pulse is applied for pre-selection to achieve high-fidelity qubit initialization. Following state preparation, the subsequent $M_2$ pulse is implemented with various pulse profiles to evaluate its ring-up contribution to MIST. Finally, a weak $M_3$ pulse is introduced to characterize both the ring-up and ring-down contributions originating from $M_2$.}
\label{fig:FIG2}
\end{figure}

By utilizing qubit devices with a relatively low $E_J/E_C$ ratio, 
the resulting charge dispersion $\delta f_{01}$ associated with the two charge-parity states \cite{Serniak2018_PRApp} is sufficient to precisely resolve and control $n_g$ through an external voltage bias, where $\delta f_{01}(n_g=0)\approx 0.1$\,MHz and $0.032$\,MHz for Q1 and Q2, respectively. Using Ramsey interferometry, two distinct qubit frequencies, which manifest as a function of the applied offset charge, are clearly resolved at a specific $n_g$ \cite{Serniak2018_PRApp}, see Fig.~\ref{fig:FIG2}(b). However, the stochastic drift and abrupt jumps of the offset charge $n_g$, occurring on timescales of several minutes \cite{Serniak2018_PRApp,Serniak2018_PRL,Christensen2019_PRB}, impose a fundamental time constraint on the data acquisition. To mitigate these low-frequency fluctuations, we implement a measurement protocol involving a comprehensive sampling of the discretized parameter manifold within each acquisition cycle, with a 10-second Ramsey-based $n_g$ calibration (Fig.~\ref{fig:FIG2}(b)) conducted every 50 seconds to maintain the target $n_g$. Specifically, the parameter space mapping is conducted with strictly one readout shot per coordinate. Through the integrated execution of parameter-space sampling and periodic recalibration cycles, we ensure that temporal charge instabilities are stochastically averaged across the entire dataset, thereby preserving the consistency of the $n_g$-dependent readout characteristics.

\begin{figure*}[t]  
\centering
\includegraphics[width=\textwidth]{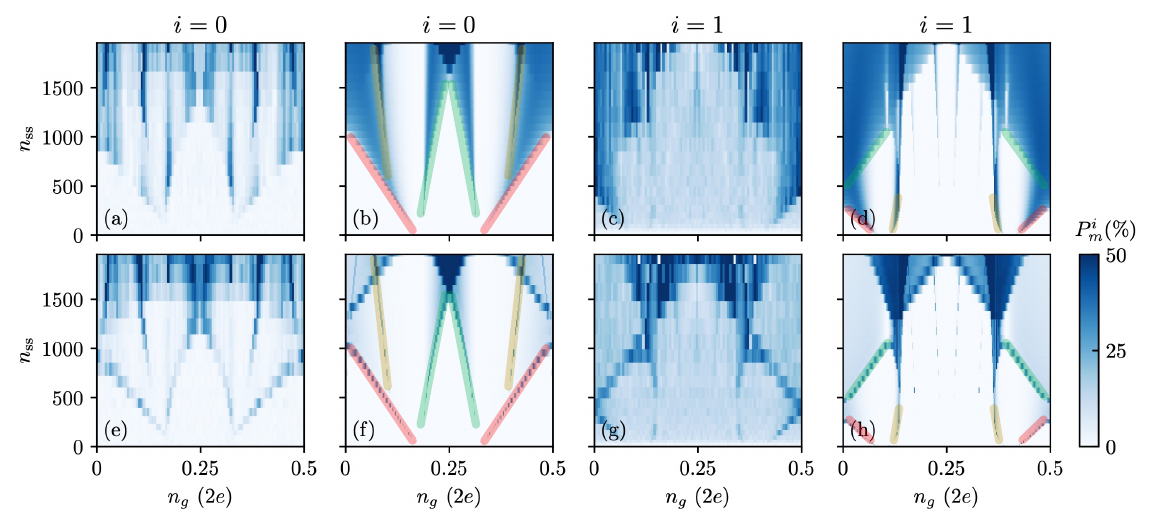} 
\caption{MIST structures associated with $M_2$ and their suppression via pulse shaping for initial qubit states $|i\rangle$. (a),(c) Measured and (b),(d) simulated MIST structures versus $n_g$ and $n_{\text{ss}}$, characterized by $P^i_m$ for a square pulse. The experimental data are benchmarked against numerical simulations via Eq.~(\ref{outlier_cal}). (e),(g) Measured and (f),(h) simulated suppression of MIST features, achieved via a calibrated two-step pulse scheme with $\tau_\uparrow = 100$\,ns and a state-independent amplitude $\varepsilon_{\uparrow}$ determined by Eq.~\eqref{cal_two_step}. The red, green, and brown semi-transparent guide lines overlaying panels (b),(f) are applied for the qubit prepared in $\vert0\rangle$ to trace the transitions from $B_0$ to branches $B_6$, $B_8$, and $B_9$, respectively. Similarly, the same color sequence is applied in panels (d),(h) for the qubit prepared in $\vert1\rangle$ to trace the transitions from $B_1$ to branches $B_5$, $B_6$, and $B_7$.}
\label{fig3}
\end{figure*}

Under the specified $n_g$, the measurement sequence depicted in Fig.~\ref{fig:FIG2}(c) comprises three distinct segments: $M_1$, $M_2$, and $M_3$. While $M_1$ and $M_3$ are implemented as identical weak QND readout pulses, $M_2$ serves as the primary pulse under characterization. All three pulses are applied at the frequency $\widetilde{\omega}_r$. Specifically, the pre-selection pulse $M_1$ facilitates high-fidelity qubit initialization. Then, the qubit is prepared in either the $|0\rangle$ or $|1\rangle$ state prior to the application of $M_2$. We utilize $M_2$ to assess the SNR and ring-up MIST across various pulse profiles, while a subsequent $M_3$ quantifies the MIST contributed by both the $M_2$ ring-up and ring-down. For each readout, we introduce an idle time to ensure the complete depletion of resonator photons prior to the subsequent operation. The following sections present a comparative analysis of $P_{m}$ and SNR for conventional square pulses and the calibrated two-step and three-step pulse profiles used in $M_2$.

\section{Experimental Results}

An ideal QND measurement requires state transitions to be suppressed not only during the measurement window but also after it concludes. However, these two regimes are physically nonequivalent. Based on the readout scheme detailed in Sec.~\ref{sec:Exp_meth}, a transition occurring within the readout integration window including the resonator ring-up and its steady-state manifests directly in the $M_2$ signal and can therefore be identified and excluded via post-selection. Conversely, a transition during the ring-down leaves the readout signal intact. While the current qubit state is assigned correctly, it leaves the qubit in an unintended state, silently violating the QND condition for subsequent operations. These ring-down transitions are intrinsically difficult to detect from $M_2$ because standard weighting schemes weight the steady-state portion of the signal to maximize the SNR, so the ring-down carries too small a fraction of the integrated signal to resolve a transition reliably. To overcome this limitation, we implement a weak post-measurement probe, $M_3$, to detect these ring-down induced transitions. Consequently, we analyze the two regimes in turn. In Sec.~\ref{sec:during}, we examine the transitions within window $M_2$ subject to post-selection, while in Sec.~\ref{sec:after}, we detail the ring-down dynamics accessed via $M_3$.

\begin{figure*}[t]
\centering
\includegraphics[width=\textwidth]{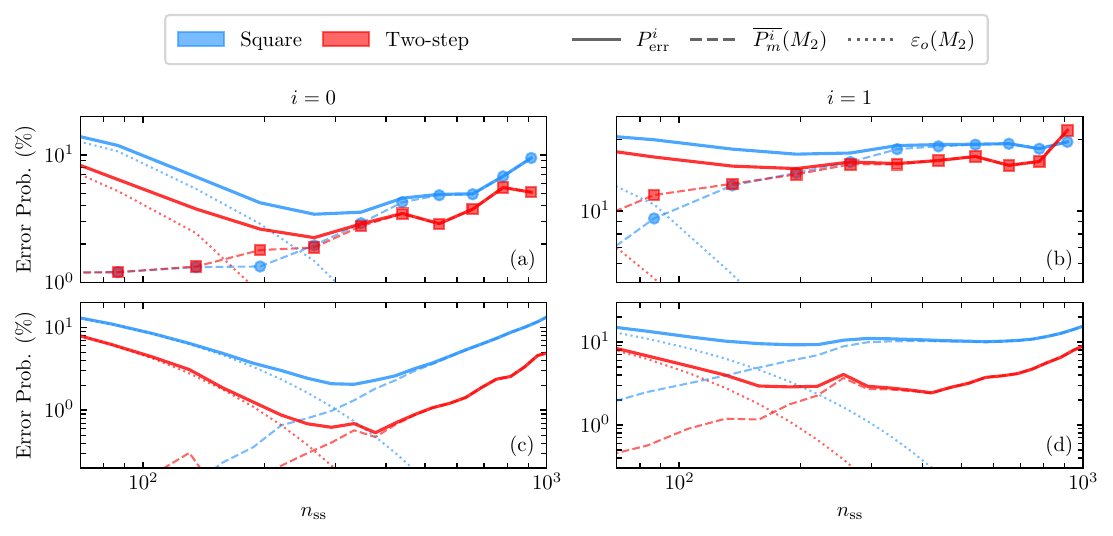} 
\caption{Readout error probabilities for initial qubit states $\vert i\rangle$ as a function of $n_{\mathrm{ss}}$, extracted from the data in Fig.~\ref{fig3}. Blue and red denote the square and two-step pulses, respectively. Throughout all panels, solid, dashed, and dotted curves represent the total readout error probability $P_{\mathrm{err}}^i$, the $n_g$-averaged transition probability $\overline{P^i_{m}}(M_2)$, and the overlap error $\varepsilon_{o}(M_2)$, respectively. (a),(b) Experimental data. The measured $\overline{P^i_{m}}(M_2)$ for the square and two-step pulses are indicated by dashed curves with circles (blue) and squares (red), respectively. (c),(d) Corresponding numerical simulations. }
\label{fig4}
\end{figure*}

\subsection{Mitigation of MIST during readout}\label{sec:during}

Because the readout signal is integrated only over the window $\tau_\uparrow+\tau_{\text{ss}}$ and excludes the resonator ring-down (as noted above and in Fig.~\ref{fig:FIG2}(c)), the two-step and three-step pulses, which differ only in their ring-down segment, produce identical $M_2$ results. We therefore restrict the analysis of the readout window to the square and calibrated two-step profiles. In this section, we first map the dependence of MIST on $n_g$ and $n_{\text{ss}}$ for both profiles. With the readout frequency set to $\omega_d = \widetilde{\omega}_r$, we calibrate the temporal evolution of the resonator photon population for each pulse via ac-Stark shift measurements (detailed in Appendix~\ref{appendix_Photon_number_calibration}). Following the protocol outlined in Sec.~\ref{sec:Exp_meth}, the experimental sequence utilizes weak square pulses $M_1$ and $M_3$ with the durations of $8\,\mu\text{s}$ and photon numbers of approximately $n_{\text{ss}}= 92$ (157) for Q1 (Q2), respectively. The $M_2$ pulse duration is fixed at $4\,\mu\text{s}$ for both devices. An idle time of $8\,\mu$s is implemented for each readout to ensure complete resonator photon initialization, especially following $M_2$ pulses with higher average photon numbers. Experimentally, the values of $P^i_{m}$ are determined by setting a threshold at three standard deviations ($3\sigma$) from the center of the Gaussian distribution corresponding to the qubit state $|i\rangle$, yielding the relation
\begin{equation}
P_{m}^{i} = 1 - P(i\mid i),
\end{equation}
where $P(i\mid i)$ is the conditional probability of measuring state $|i\rangle$ given the prepared state $|i\rangle$.

Figure~\ref{fig3} presents the measured $P^i_{m}$ values for Q1 alongside numerical simulations to systematically investigate the behavior of MIST under both square (Figs.~\ref{fig3}(a)--(d)) and calibrated two-step (Figs.~\ref{fig3}(e)--(h)) pulse profiles. In these panels, multiple band structures characterized by high transition probabilities are clearly visible, indicating the presence of MIST. The background probability is higher in the experiment than that in the simulation, an offset we attribute to decoherence, which is not accounted for in our model.  To identify the underlying physics of these features, we perform a numerical Floquet branch analysis. For instance, the red, green, and brown semi-transparent guide lines in Figs.~\ref{fig3}(b) and (f) trace the transitions between $B_0$ and the branches $B_6$, $B_8$, and $B_9$, respectively; similarly, those in Figs.~\ref{fig3}(d) and (h) follow the same color sequence to map the transitions between $B_1$ and the branches $B_5$, $B_6$, and $B_7$, respectively. As demonstrated in Fig.~\ref{fig1}, MIST occurs when the dynamic trajectory of $n(t)$ for a given $n_g$ traverses these labeled resonance structures. Here, the $n(t)$ trajectory generated by the $M_2$ pulse during the ring-up process toward a specific $n_{\text{ss}}$ is utilized to calculate $P^i_m$ from the pulse start time $t_1$ to its end time via Eq.~(\ref{outlier_cal}), yielding the total transition probability $P^i_m(t_1, t_1+\tau_\uparrow+\tau_{\text{ss}})$ plotted in the numerical simulations of Fig.~\ref{fig3}. Crucially, aside from the highly hybridized regions, Figs.~\ref{fig3}(e)--(h) demonstrate that these prominent MIST features are suppressed across the entire $n_g$ and $n_{\text{ss}}$ plane under the two-step pulse scheme. This mitigation is achieved using a state-independent amplitude $\varepsilon_{\uparrow}$ given by Eq.~\eqref{cal_two_step}, with parameters $\tau_\uparrow = 100$\,ns and $\tau_\uparrow+\tau_{\text{ss}}=4\,\mu$s. The higher background probabilities for the prepared $\vert1\rangle$ state in Figs.~\ref{fig3}(c) and (g) relative to the $\vert0\rangle$ state in Figs.~\ref{fig3}(a) and (e) stem primarily from statistical uncertainties and an estimated $M_2$ relaxation error of $\tau/2T_1 \sim 0.68\%$, where $\tau = 4\,\mu\text{s}$. As anticipated in Sec.~\ref{section_fast_load}, the enhanced ring-up rate $\Gamma_{i,r}(n_g)$ effectively mitigates $P_{m}$ during the fast-loading stage, validating our optimization strategy. In contrast to the square-pulse scenario, these labeled branch transitions are clearly resolved in both the experimental and simulated data uniquely under this two-step framework.

Note that the $P^i_{m}$ profiles exhibit a reflection symmetry with respect to $n_g = 0.25$. This symmetry arises because the rapid parity switching occurs on a timescale much shorter than our measurement time, leading to equally populated parity states and an effective symmetry in $P^i_{m}$. Exploiting this symmetry, the experimental data were measured solely within $n_g \in [0, 0.25]$ and mirrored onto the $[0.25, 0.5]$ range. Consistently, the simulation accounts for this effective parity ensemble by averaging the numerical results at $n_g$ and $0.5 - n_g$ over the $[0, 0.5]$ interval. For conciseness, the corresponding results for Q2 are provided in the Supplementary Material.

\begin{figure*}[t]
\centering
\includegraphics[width=\textwidth]{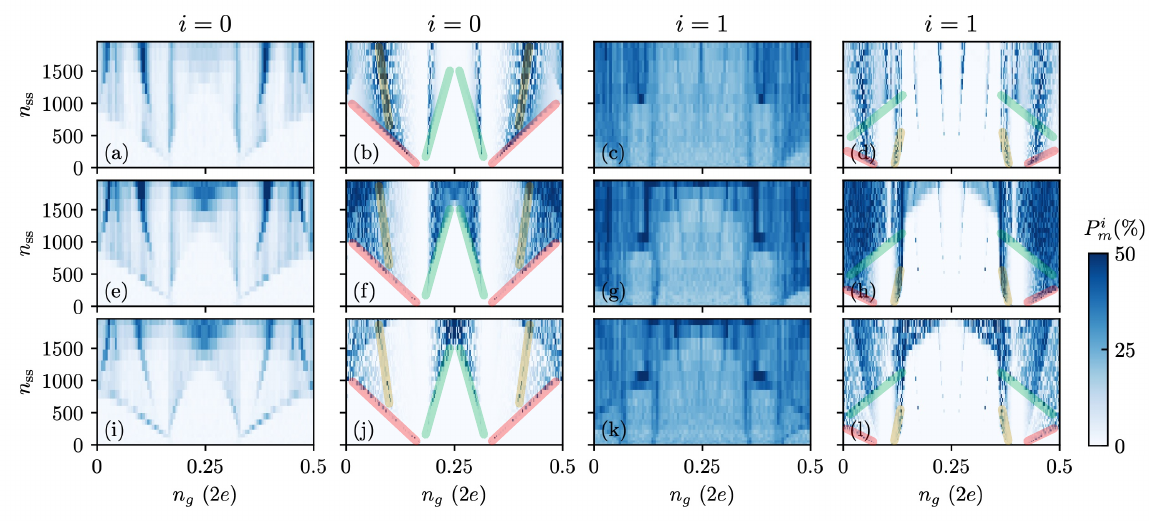} 
\caption{MIST structures associated with $M_3$ for initial qubit states $\vert i\rangle$. The measured MIST, characterized by $P_{m}^{i}$, is depicted as a function of $n_g$ and $n_{\text{ss}}$ under various $M_2$ pulse shapes, where (a),(c) denote the square pulse, (e),(g) the two-step pulse, and (i),(k) the three-step pulse. The corresponding simulated results are presented in (b),(d); (f),(h); and (j),(l), respectively. The colored semi-transparent guide lines correspond to the branch transitions defined in Fig.~\ref{fig3}. }
\label{fig5}
\end{figure*}

To systematically evaluate the trade-off between the readout SNR and MIST, we investigate the readout total error probability
\begin{equation}
P_{\text{err}}^i=\overline{P^i_{m}}(M_2)+\varepsilon_{o}(M_2),
\end{equation} 
which comprises the $n_g$-averaged transition probability $\overline{P^i_{m}}$ and the overlap error $\varepsilon_{o}$ calculated from the SNR as defined in Eq.~\eqref{OE}, both of which are extracted directly from $M_2$ pulse. Here, the SNR is experimentally estimated from single-shot statistics via a double-Gaussian analysis as $\text{SNR}=D/\sigma$, where $D$ represents the separation between the centers of the $\vert0\rangle$ and $\vert1\rangle$ state distributions, and $\sigma$ denotes their standard deviation on the IQ plane. 

The results extracted from Fig.~\ref{fig3} are summarized in Fig.~\ref{fig4}; we exclude the low-SNR ($n_{\text{ss}} < 70$) and highly hybridized ($n_{\text{ss}} > 1000$) regimes, where an accurate quantification of $P^i_{m}$ becomes infeasible. To incorporate $\varepsilon_{o}$ into the simulation framework, the experimental SNR is subsequently extrapolated using a linear fit against $\sqrt{n_{\text{ss}}}$.  In Fig.~\ref{fig4}, panels (a) and (b) present the experimental measurements, while panels (c) and (d) display the corresponding numerical simulations. Notably, both the experimental and simulated data exhibit a lower $P^i_{\text{err}}$ in the dynamic regime for the two-step pulse (red solid curves) relative to the square pulse (blue solid curves), and the experimental trends show good agreement with our numerical simulations. Consequently, this reduction in $P^i_{\text{err}}$ demonstrates that both enhancing readout speed and mitigating MIST are achieved by the fast-load pulse. The noticeable discrepancy between Figs.~\ref{fig4}(b) and (d) results from the same relaxation issue discussed previously in the context of Figs.~\ref{fig3}. Although we successfully characterize the $n_g$-averaged readout errors originating from MIST, in realistic scenarios, the temporal drift of $n_g$ is inevitable. Therefore, $n_g$-averaged characterization provides a more robust and conservative metric that faithfully reflects the long-term overall readout fidelity.

\subsection{Mitigation of MIST after readout}\label{sec:after}

We now use $M_3$ to probe the post-measurement state. The $M_2$ results detailed in Sec.~\ref{sec:during} are insensitive to the ring-down and therefore identical for the two-step and three-step pulses. In contrast, the final qubit state reflects the full pulse evolution, including the ring-down. Hence, the square, two-step, and three-step profiles can leave the qubit in distinct post-measurement states.

Figure~\ref{fig5} compares the $M_3$ responses under square (panels (a) to (d)), two-step (panels (e) to (h)), and three-step (panels (i) to (l)) $M_2$ pulse profiles. Here, the total integration window for all configurations is set to $4\,\mu\text{s}$. The multi-step configurations utilize state-independent treatments with $\tau_\uparrow = \tau_\downarrow = 100$\,ns and $\tau_\uparrow + \tau_{\text{ss}} = 4\,\mu\text{s}$, as detailed in Appendix~\ref{appendix_three_step}. Unlike the $M_2$ results in Fig.~\ref{fig3}, where the two-step pulse exhibits lower MIST than the square pulse, $M_3$ displays the converse behavior due to the critical impact of the resonator ring-down process. By employing the three-step pulse to actively accelerate the ring-down, we observe a mitigation of MIST relative to the two-step scheme. However, the performance advantage predicted for the three-step pulse over the square pulse in Fig.~\ref{fig1} does not manifest experimentally in Fig.~\ref{fig5} across the entire parameter space; in fact, the two-step pulse exhibits the most severe MIST.  This behavioral contrast becomes highly pronounced when comparing the reference square-pulse results in Figs.~\ref{fig5}(a) and (c) against the significantly broadened transition bands in Figs.~\ref{fig5}(e) and (g). The higher background probabilities for the prepared $\vert 1\rangle$ state in Figs.~\ref{fig5}(c), (g), and (k) stem primarily from the relaxation error occurring from state preparation through the end of $M_3$, corresponding to the pulse sequence depicted in Fig.~\ref{fig:FIG2}(c).

\begin{figure*}[t]
\centering
\includegraphics[width=\textwidth]{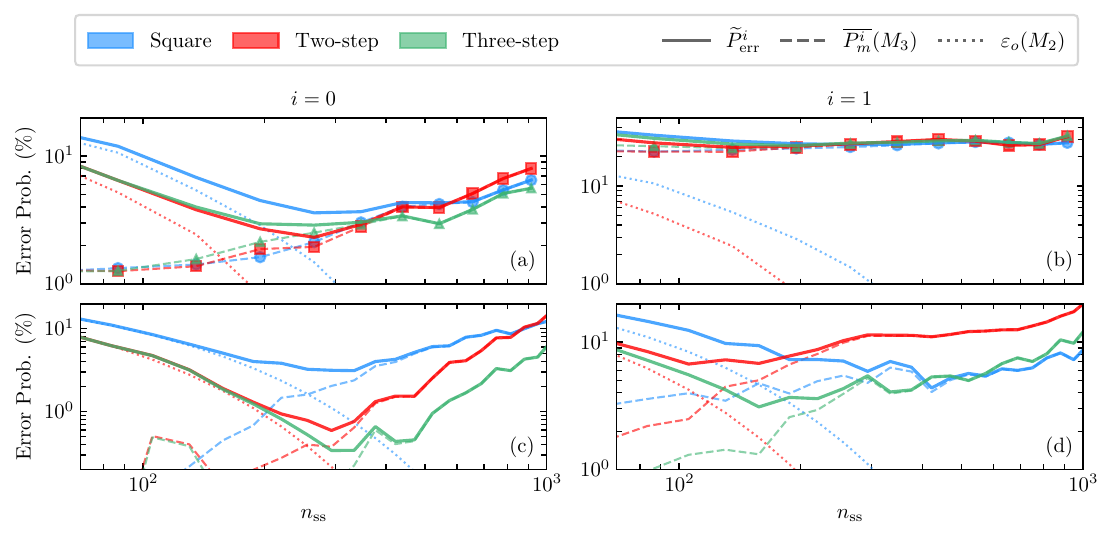} 
\caption{Post-readout error probabilities for initial qubit states $\vert i\rangle$ as a function of $n_{\mathrm{ss}}$, extracted from the data in Fig.~\ref{fig5}. Blue, red, and green denote the square, two-step, and three-step pulses, respectively. Throughout all panels, solid, dashed, and dotted curves denote the post-readout total error probability $\widetilde{P}_{\mathrm{err}}^i$, $n_g$-averaged transition probability $\overline{P^i_{m}}(M_3)$, and the overlap error $\varepsilon_{o}(M_2)$, respectively. (a),(b) Experimental data. The measured $\overline{P^i_{m}}(M_3)$ for the square, two-step, and three-step pulses are indicated by dashed curves with circles (blue), squares (red), and triangles (green), respectively. (c),(d) Corresponding simulated data.}
\label{fig6}
\end{figure*}

These observations can be understood in terms of the photon trajectory dynamics, which stem from LZS interferometry. During the ring-up process, the system undergoes a coherent probability splitting into adiabatic and diabatic branches, with the latter physically preserving the original bare state. As the system evolves prior to the second traversal, a relative dynamical phase accumulates between these two branches. Consequently, upon the second traversal of the avoided crossing during the ring-down process, these branches recombine and interfere, unavoidably affecting the final state distribution via this accumulated phase. Notably, in either the extremely diabatic or adiabatic limits, the high degree of symmetry in the photon trajectory robustly preserves the initial bare state. This highlights the key advantage of both the square and three-step pulses, which feature a significantly more symmetric $\Gamma_{i,r}(n_g)$ profile between the ring-up and ring-down processes than that of the two-step pulse, thereby explaining their better performance. This trajectory symmetry directly accounts for the suppression of severe post-readout MIST in the three-step configuration (Figs.~\ref{fig5}(i) and (k)) relative to the asymmetric two-step pulse (Figs.~\ref{fig5}(e) and (g)). In principle, under an ideal symmetric photon trajectory and an appropriate accumulated dynamic phase, both slow and fast transients can preserve the initial bare state. However, the former requires a longer readout duration to achieve the same SNR, rendering the preservation more susceptible to qubit decoherence. Because the pulse duration was fixed in this work, the accumulated dynamic phase remained constant, leaving a systematic exploration of LZS interference beyond the scope of this study.

Conceptually, the simulations characterizing $P^i_m$ from $M_3$ should follow the same treatment as $M_2$ in Sec.~\ref{sec:during}; however, for simplicity, the $M_3$ pulse is assumed to be strictly adiabatic, allowing us to utilize the instantaneous transition probability at the initiation of $M_3$. Since the relaxation time is significantly longer than both the duration of $M_3$ and the idle time between $M_2$ and $M_3$, the branch probabilities remain unchanged after the $M_2$ ring-down. Therefore, this allows us to isolate the pure effect of $M_2$. The numerical simulations exhibit good agreement with the experimental data, particularly within the strongly hybridized regions of the main MIST structures.  Specifically, the simulated profiles for the square (Figs.~\ref{fig5}(b) and (d)), two-step (Figs.~\ref{fig5}(f) and (h)), and three-step (Figs.~\ref{fig5}(j) and (l)) pulses are in good agreement with their experimental counterparts. The same colored semi-transparent guide lines in Fig.~\ref{fig3} are applied to the branch structures in Fig.~\ref{fig5}. It is worth noting that any given photon trajectory results in the same MIST onset as a function of $n_g$ and $n_{\text{ss}}$, which is intrinsically determined by the system characteristics. Conversely, the width of these band structures and the severity of the MIST are dynamically governed by $n(t)$. The variations shown in the simulations in Fig.~\ref{fig5} do not represent numerical artifacts or random noise, but rather originate from the coherent quantum interference occurring upon the second traversal of the avoided crossing.

Similarly, a consistent treatment is applied by defining the post-readout total error probability as
\begin{equation}
\widetilde{P}_{\text{err}}^i = \overline{P^i_{m}}(M_3) + \varepsilon_{o}(M_2),
\end{equation}
which is composed of the $n_g$-averaged transition probability $\overline{P^i_{m}}$ from $M_3$ and $\varepsilon_{o}$ from $M_2$. Importantly, the portion of $\widetilde{P}_{\text{err}}^i$ contributed by LZS remains immune to post-selection because the $M_2$ measurement is inherently insensitive to the ring-down transitions, which are resolved exclusively by the subsequent $M_3$ probe, while $\varepsilon_{o}$ represents an intrinsic residual error arising from finite state distinguishability. Physically, the term $\overline{P^i_{m}}(M_3)$ fundamentally constrains the fidelity of subsequent quantum information processing. In Fig.~\ref{fig6}, we compare $\widetilde{P}_{\text{err}}^i$ among the square (blue solid curves), two-step (red solid curves), and three-step (green solid curves) pulses by using the data extracted from Fig.~\ref{fig5}.

When incorporating the effect of the SNR, the error landscape shifts significantly compared to the pure transition trends observed in Fig.~\ref{fig5}, where the two-step scheme represents the worst-case scenario. Specifically, the simulation results in Figs.~\ref{fig6}(c) and (d) reveal that the three-step pulse serves as the optimal configuration, whereas the square pulse exhibits the worst performance across the dynamic range. Qualitatively, our experimental results for the qubit prepared in $\vert0\rangle$ in Fig.~\ref{fig6}(a) partially corroborate these predictions. While the square pulse remains the least effective scheme as predicted by simulations, the performance of the two-step and three-step schemes is comparable across the entire dynamic range. Moreover, Figs.~\ref{fig6}(b) and (d) display a more pronounced discrepancy between the experimental and simulated results for the $\vert1\rangle$ state than the $P_{\text{err}}^i$ case previously shown in Figs.~\ref{fig4}(b) and (d). This deviation primarily arises from a combination of the non-negligible MIST contribution induced during the $M_3$ process itself and a significant relaxation error of $\tau/2T_1 \sim 3.4\%$. The latter stems from an extended total duration of $\tau = 20\,\mu\text{s}$ that accounts for the combined durations of the $M_2$ pulse, the $M_3$ pulse, and their intermediate idle time. The $M_3$-associated MIST as a function of $n_g$ can be further anticipated from the $M_2$ profiles in Figs.~\ref{fig3}(b) and (d) given $n_{\text{ss}} = 92$.

Although we use a device with a relatively narrow resonator linewidth ($\kappa/2\pi \approx 0.392\,\text{MHz}$) to demonstrate these mechanisms, a larger $\kappa$ would inherently enable faster readout and enhanced measurement performance, thereby yielding closer agreement between experimental observations and numerical simulations. While the pulse-shape advantages are experimentally obscured for $\vert1\rangle$, the unambiguous validation for $\vert0\rangle$ complemented by idealized simulations firmly confirms that the three-step pulse scheme provides a highly reliable post-readout state with minimized total error. Because achieving high QND fidelity is a strict prerequisite for executing subsequent conditional quantum gates, this scheme demonstrates superior performance for fast feedback operations compared to alternative pulse shapes. Ultimately, this multi-step engineering approach successfully suppresses post-readout backaction without compromising the readout speed.

\section{Conclusion}
In this work, we have experimentally investigated the trade-off between the $n_g$-dependent MIST and the enhancement of readout SNR under fast-load and fast-clear pulse operations. To evaluate the MIST stemming from different pulse profiles without loss of generality, we constructed simple state-independent multi-step pulses to achieve rapid photon ring-up and ring-down with suppressed transient photon overshoots. By utilizing Floquet branch analysis within a semiclassical model, we directly compared the time-dependent dressed-state hybridization among various pulse shapes, successfully identifying the underlying resonant band structures. Through a consecutive readout pulse protocol, this framework enables us to characterize both the independent ring-up contribution and the combined ring-up and ring-down backaction. Moreover, we investigated the dependence of the $n_g$-averaged total error probabilities for both readout and post-readout stages on the resonator photon number. Focusing on the ring-up dynamics reveals that a higher diabaticity, which shortens the resonator transient time, inherently suppresses readout errors. This demonstrates that fast readout can be realized without compromising MIST mitigation.  Additionally, preserving the post-readout quantum state requires a highly symmetric pulse trajectory during the ring-down transient to minimize residual backaction.

Within the readout window, the fast-load pulses reduce the total readout error relative to the square pulse in both experiment and simulation. Our numerical simulations further demonstrate that the three-step pulse minimizes the post-readout total error across the dynamic range, and the experimental data are consistent with this prediction, unambiguously resolving its advantage over the square pulse for the qubit prepared in $\vert 0\rangle$. The three-step scheme therefore matches the fast-load performance during readout while additionally preserving the post-measurement state, making it the only profile examined that is not compromised at either stage across the entire $n_g$ range. This capability enables rapid feedback control that facilitates a more precise active reset required for long-$T_1$ devices. Consequently, this pulse-shaping approach offers a practical route to enhancing readout fidelity and preserving the QND character of dispersive readout without requiring additional hardware. It is especially relevant to applications requiring repeated mid-circuit measurements, such as quantum error correction.

\section*{Data availability}
The data that support the findings of this study are available from the corresponding author upon reasonable request.

\begin{acknowledgments}
We thank for the support from the QC-Test at the Research Center for Critical Issues (RCCI), Academia Sinica, Taiwan.  This work was supported by the National Science and Technology Council (NSTC) in Taiwan via Grants No. NSTC 113-2119-M-001-008, NSTC 114-2119-M-001-004, and from Academia Sinica via Grants No. AS-GCP-112-M01, AS-GCS-114-M04, AS-KPQ-111-TQRB.
\end{acknowledgments}

\appendix
\section{Experiment setup}\label{Experiment Setup}
\begin{figure} 
\centering
\includegraphics[width=\columnwidth]{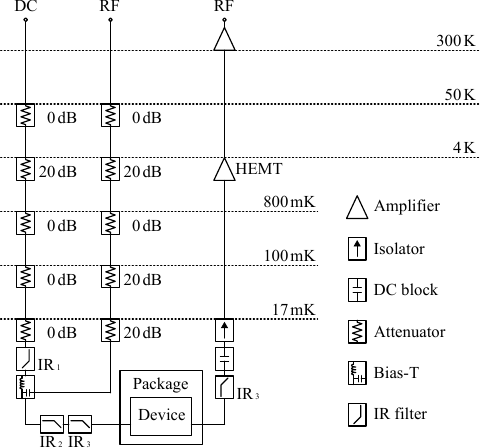} 
\caption{Schematic of the experimental wiring and measurement setup.}
\label{AP_D0}
\end{figure}

Here, we detail the experimental setup of the signal chain. To suppress stray infrared (IR) radiation propagating through the coaxial lines, a comprehensive filtering configuration is implemented along each signal path, as schematically illustrated in Fig.~\ref{AP_D0}. A low-cutoff Eccosorb filter (Quantum Microwave, QMC-CRYOIRF-001MF-S, CR-124 epoxy, labeled as $\text{IR}_{1}$) is implemented on the DC bias line. For the RF input drive line, the circuit incorporates a multi-stage filtering scheme. An absorptive Eccosorb filter (Bluefors, Low-loss Bulkhead IR filter, labeled as $\text{IR}_{2}$) is connected in series with a non-magnetic High-Energy Radiation-Drain (HERD) filter (Sweden Quantum, HERD-2FBLK, labeled as $\text{IR}_{3}$). Conversely, the output line contains only a single $\text{IR}_{3}$ filter to optimize the effective system noise temperature of the amplification chain. This asymmetric configuration is strategically dictated by the ultra-low passband insertion loss of the HERD filter, thereby preserving the readout SNR.  

The device is housed in a nested shielding configuration consisting of an innermost copper shield, an aluminum shield equipped with a light-tight indium seal, and an outermost mu-metal shield. To mitigate stray infrared radiation, the interior surfaces of both the copper and aluminum shields, as well as the mixing-chamber shield, are coated with an IR-absorber. The details for the package are provided in Ref.~\cite{Package_arxiv}. 

\begin{figure}
\centering
\includegraphics[width=\columnwidth]{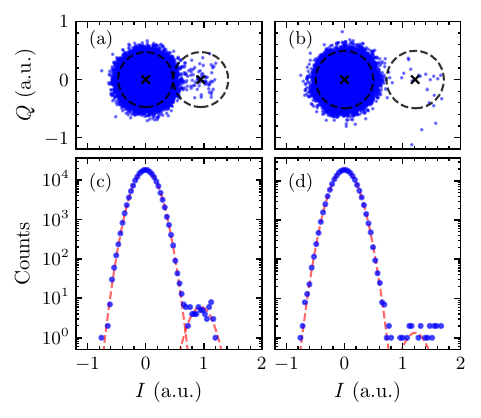}
\caption{Single-shot qubit readout distributions with $2\times10^5$ shots initialized in $\lvert0\rangle$. (a),(b) Readout signals in the in-phase and quadrature (IQ) plane for Q1 and Q2, respectively, with both qubits prepared in $|0\rangle$. The black dashed circles denote the $3\sigma$ boundaries of the distributions. (c),(d) The corresponding histograms (blue dots) and the associated double-Gaussian fits (red curves).}
\label{fig:AP_D1}
\end{figure}

As depicted in Fig.~\ref{fig:AP_D1}, the effective qubit temperatures are determined to be $T_q = 18.46$\,mK and $16.86$\,mK, which correspond to ultra-low excited-state populations of $P_1 \approx 0.033\%$ and $0.007\%$, respectively, closely approaching the mixing chamber base temperature $T_{\rm{MX}}\sim 17$\,mK. These values are extracted from $2\times10^5$ single-shot measurements performed in the absence of qubit control pulses, with the data analyzed using a double-Gaussian fit. Such low effective temperatures successfully demonstrate the high effectiveness of our shielding, packaging design and the IR filtering.

\section{Analytical expression of resonator fields}\label{appendix_io_theory}
We model a notch-type readout resonator coupled to both a transmon qubit and a transmission line with coupling strengths $g$ and $\kappa_c$, respectively. The transmitted and reflected light fields are denoted as $b^{\text{in,t}}$ and $b^{\text{in,r}}$ at the input port, and as $b^{\text{out,t}}$ and $b^{\text{out,r}}$ at the output port.  In the dispersive regime ($|\Delta_{qr}|=|\omega_{01}-\omega_r|\gg g$), the resonator frequency is pulled by a state-dependent shift $\chi$. The resonator fields are characterized by coherent state amplitudes $\alpha_{\pm}(t)$, where the minus and plus signs $\pm$ correspond to $|0\rangle$ and $|1\rangle$, respectively. The readout signal is applied as $b^{\text{in,t}}$ at the input port and measured as $b^{\text{out,t}}$ at the output port, assuming zero reflection ($b^{\text{out,r}}=0$). Following the standard input-output formalism \cite{Gardiner1985,Clerk2010}, the time evolution of resonator fields in the rotating frame of driving frequency $\omega_d$ can be described by the equations
\begin{equation}
\dot\alpha_{\pm}=-i(\Delta_r\pm\chi)\alpha_{\pm}-\frac{\kappa}{2}\alpha_{\pm}-\sqrt{\frac{\kappa_c}{2}}b^{\text{in,t}},
\end{equation}
where $\kappa=\kappa_i+\kappa_c$ is the total decay rate incorporating the intrinsic loss rate $\kappa_i$, and $\Delta_r = \widetilde{\omega}_r - \omega_d$. $\widetilde{\omega}_r$ represents the average frequency of the dispersively shifted resonator frequencies corresponding to the qubit states.

The amplitude of the readout field is given by $b^{\text{in,t}}=\sqrt{\frac{P_{\text{in}}(t)}{\hbar\omega_d}}$ with an input power $P_{\text{in}}(t)$. Under a constant readout drive $b^{\text{in,t}} = \varepsilon$, the general solution for the resonator fields corresponding to each qubit state is given by
\begin{equation}
\alpha_{\pm}(t)=\left(\alpha_{\pm}(0)-\mathcal{A}_{\pm}\right)e^{-[\kappa/2+i(\Delta_{r}\pm\chi)]t}+\mathcal{A}_{\pm},
\label{AP_pho_field}
\end{equation}
where the steady-state amplitude $\mathcal{A}_{\pm}$ is expressed as
\begin{equation}
\mathcal{A}_{\pm}(\varepsilon) = \frac{-\varepsilon \sqrt{\kappa_c/2}}{\kappa/2 + i(\Delta_r \pm \chi)}.
\label{AP_ss_field}
\end{equation}
Following the input-output  conditions, the transmitted output fields 
for each qubit state are expressed as $b^{\text{out,t}}_{\pm}(t) = \varepsilon + \sqrt{\frac{\kappa_c}{2}} \alpha_{\pm}(t)$. 
The signal contrast between the two qubit states is defined as
\begin{equation}
S(t) \equiv \sqrt{\frac{\kappa_c}{2}} \left| \alpha_{+}(t) - \alpha_{-}(t) \right|.
\end{equation}
Accordingly, the qubit-state SNR is defined as the ratio of the integrated signal contrast to the integrated noise fluctuations over an integration window of duration $\tau$ starting from $t_i$
\begin{equation}
\text{SNR} = \sqrt{\frac{\eta}{\tau}} \int_{t_{i}}^{t_{i}+\tau} S(t) dt,
\label{SNR}
\end{equation}
where $\eta = 1/N_{s} = \hbar\omega_{d}/k_{B}T_{s}$ represents the quantum efficiency, with $T_{s}$ being the effective system noise temperature and 
$N_{s}$ the corresponding noise photon number referred to the sample stage. Finally, the overlap error rate $\varepsilon_{o}$ is determined by the SNR \cite{Gambetta:2007pra, Wilkinson:2024prxquantum}
\begin{equation}
\varepsilon_{o} = \frac{1}{2} \left[ 1 - \text{erf} \left( \frac{\text{SNR}}{\sqrt{8}} \right) \right].
\label{OE}
\end{equation}

\section{Fast-load and fast-clear readout via three-step pulses}\label{appendix_three_step}

Following the framework in Appendix~\ref{appendix_io_theory}, we introduce a three-step pulse consisting of three functionally distinct segments defined by their respective drive amplitudes $\varepsilon$ and durations $\tau$: the initial ring-up segment $\{\varepsilon_\uparrow, \tau_\uparrow\}$ designed to rapidly populate the resonator, the steady-state segment $\{\varepsilon_{\text{ss}}, \tau_{\text{ss}}\}$ to sustain the readout at target photon number, and the active ring-down segment $\{\varepsilon_\downarrow, \tau_\downarrow\}$ to accelerate the depletion of resonator photons. To facilitate fast loading and fast clearing of the readout resonator, the pulse amplitudes are determined by imposing the conditions $\alpha_{\pm}(\tau_\uparrow) = \mathcal{A}_{\pm}(\varepsilon_{\text{ss}})$ and $\alpha_{\pm}(\tau_\uparrow + \tau_{\text{ss}} + \tau_\downarrow) = 0$. Therefore, the state-dependent amplitudes $\varepsilon_{\uparrow,\pm}$ and $\varepsilon_{\downarrow,\pm}$ are uniquely determined by the steady-state drive $\varepsilon_{\text{ss}}$ and the segment durations as
\begin{equation}
\varepsilon_{\uparrow,\pm} = \frac{\varepsilon_{\text{ss}}}{1 - e^{-[\kappa/2 + i(\Delta_r \pm \chi)] \tau_\uparrow}}
\end{equation}
and
\begin{equation}
\varepsilon_{\downarrow,\pm} = \frac{-\varepsilon_{\text{ss}}}{e^{[\kappa/2 + i(\Delta_r \pm \chi)] \tau_\downarrow} - 1}.
\end{equation}

For a more practical implementation that ensures near-symmetric ring-up and ring-down behavior without requiring state-dependent feedback, we consider the case where $\Delta_r = 0$, $\tau_\uparrow = \tau_\downarrow$, $\left|\alpha_{\pm}(\tau_\uparrow)\right| = \left|\mathcal{A}_{\pm}(\varepsilon_{\text{ss}})\right|$ and $\left|\alpha_{\pm}(\tau_\uparrow + \tau_{\text{ss}} + \tau_\downarrow)\right| = 0$. Under these conditions, the required drive magnitudes become identical for both qubit states, allowing the use of state-independent amplitudes
\begin{equation}
\varepsilon_{\uparrow} = \left| \frac{\varepsilon_{\text{ss}}}{1 - e^{-(\kappa/2 + i\chi) \tau_\uparrow}} \right|
\label{R_up_condition}
\end{equation}
and
\begin{equation}
\varepsilon_{\downarrow} = - \left| \frac{\varepsilon_{\text{ss}}}{e^{(\kappa/2 + i\chi) \tau_\downarrow} - 1} \right|.
\label{R_down_condition}
\end{equation}
Notably, despite the lack of state-dependent phase compensation, this approach effectively suppresses the large transient photon overshoots typically observed in uncompensated ring-up pulses. Although the resonator photon population does not immediately stabilize and instead exhibits a subtle settling behavior toward $n_{\text{ss}}$, the overall transient behavior is significantly mitigated. In the meantime, the ring-down segment cannot perfectly deplete the photons to zero, resulting in a finite residual photon population at the end of the readout pulse.

\begin{figure}
\centering
\includegraphics[width=\columnwidth]{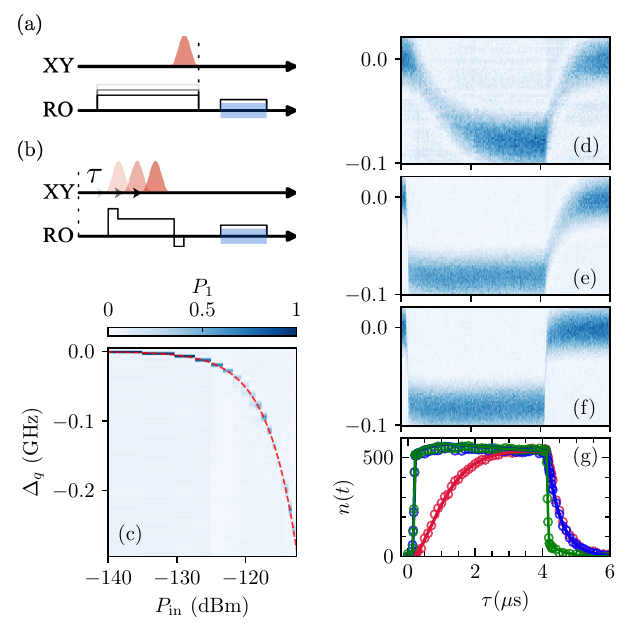}
\caption{Characterization of resonator photon dynamics. Pulse sequences for (a) ac-Stark shift calibration and (b) time-resolved $n(t)$ mapping using a variable delay $\tau$ of the $\pi$ pulse. XY (RO) labels the qubit (resonator) drive, with readout pulses shown as blue-shaded regions. (c) Qubit detuning $\Delta_q=\widetilde\omega_{01}(\varepsilon_{\text{s}}=0)-\omega_{s}$ as a function of input power $P_{\text{in}}$ used for the calibration. The red dashed curve denotes the numerical fit according to Eq.~(\ref{ac_relation}). (d)--(f) Time-resolved qubit spectroscopy of $n(t)$ mapping for square, two-step, and three-step pulses, respectively. (g) Extracted $n(t)$ (open circles) compared with input-output simulations (solid curves) for square (red), two-step (blue), and three-step (green) pulses.}
\label{fig:AP_D2}
\end{figure}

\section{Photon number calibration}
\label{appendix_Photon_number_calibration}

As illustrated in the pulse sequence in Fig.~\ref{fig:AP_D2}(a), we calibrate the resonator photon population via ac-Stark shift measurements \cite{Sank2014PhD,Sank:2016prl}. The qubit is initialized at $|0\rangle$, and its spectroscopy is performed using a 200\,ns detuned driving pulse with the frequency $\omega_{s}$ while simultaneously applying a Stark drive at $\widetilde{\omega}_r$ to populate the resonator. By extracting the Stark-shifted qubit transition frequency $\widetilde\omega_{01}$ at various Stark drive amplitudes $\varepsilon_{\text{s}}$, we calibrate the steady-state photon number $n_{\text{ss}} = \xi\varepsilon_{\text{s}}^2$ via the ac-Stark shift relation
\begin{equation}
\widetilde\omega_{01}(\varepsilon_{\text{s}}) = \omega_{01} - (2\xi\varepsilon_{\text{s}}^2 + 1)\chi,
\label{ac_relation}
\end{equation}
where $\xi$ is the calibration constant. In Fig.~\ref{fig:AP_D2}(c), the drive amplitude $\varepsilon_{\text{s}}$ is converted into the input power $P_{\text{in}}$ at the qubit device input by computing $n_{\text{ss}}$ as a function of $P_{\text{in}}$ from input-output theory via Eq.~\eqref{AP_ss_field} and matching it to the measured value of $n_{\text{ss}}$; this matching determines $\xi$ and thereby fixes the correspondence between $\varepsilon_{\text{s}}$ and $P_{\text{in}}$. To resolve the temporal dynamics of the resonator field, we perform qubit spectroscopy in Fig.~\ref{fig:AP_D2}(b) as a function of the delay between the pulse under characterization and a subsequent 40\,ns weak driving pulse. The time-dependent qubit transition frequencies are then mapped to the resonator photon evolution $n(t)$ using Eq.~(\ref{ac_relation}). We characterize the $4\,\mu$s square, two-step, and three-step pulse profiles (from Fig.~\ref{fig:AP_D2}(d)--(f)) using the conditions specified in Eqs.~(\ref{R_up_condition}) and (\ref{R_down_condition}) with $\tau_\uparrow = \tau_\downarrow = 100$\,ns and $\tau_{\text{ss}}+\tau_\uparrow=4\,\mu$s. The extracted time evolutions of the photon number for these three pulse profiles are depicted in Fig.~\ref{fig:AP_D2}(g), showing good agreement with numerical simulations based on input-output theory. The minor discrepancies may arise from pulse distortion along the readout line and uncertainties in the extracted parameters $\kappa$ and $\chi$.

\nocite{*}
\bibliography{apssamp}

\end{document}